\documentclass{v17cosmo}

\def\be{\begin{equation}}
\def\ee{\end{equation}}
\def\bea{\begin{eqnarray}}
\def\eea{\end{eqnarray}}

\newcommand{\Photo}{\includegraphics[height=35mm]{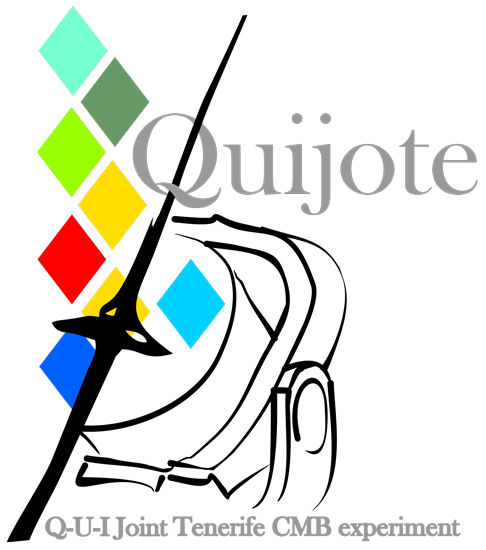}}

\begin{document}
\vspace*{4cm}
\title{THE QUIJOTE EXPERIMENT: PROSPECTS FOR CMB B-MODE POLARIZATION
  DETECTION AND FOREGROUNDS CHARACTERIZATION}
\author{ F. POIDEVIN$^{1,6}$, 
J.A. RUBINO-MARTIN$^{1,6}$,
R. GENOVA-SANTOS$^{1,6}$, 
R. REBOLO$^{1,6,7}$, 
M. AGUIAR$^{1}$,
F. GOMEZ-RENASCO$^{1}$,
F. GUIDI.$^{1,6}$,  
C. GUTIERREZ$^{1,6}$, 
R. J. HOYLAND$^{1}$,
C. LOPEZ-CARABALLO$^{1,6,8}$, 
A. ORIA CARRERAS$^{1}$,
A. E. PELAEZ-SANTOS$^{1,6}$, 
M. R. PEREZ-DE-TAORO$^{1,6}$,
B. RUIZ-GRANADOS$^{1,6}$, 
D. TRAMONTE$^{1,6}$, 
A. VEGA-MORENO$^{1}$,
T. VIERA-CURBELO$^{1}$, 
R. VIGNAGA$^{1,6}$, 
E. MARTINEZ-GONZALEZ$^{2}$,
R. B. BARREIRO$^{2}$, 
B. CASAPONSA$^{2}$, 
F. J. CASAS$^{2}$, 
J. M. DIEGO$^{2}$,
R. FERNANDEZ-COBOS$^{2}$, 
D. HERRANZ$^{2}$, 
M. LOPEZ-CANIEGO$^{2}$,
D. ORTIZ$^{2}$, 
P. VIELVA$^{2}$, 
E. ARTAL$^{3}$, 
B. AJA$^{3}$, 
J. CAGIGAS$^{3}$,
 J. L. CANO$^{3}$, 
L. DE LA FUENTE$^{3}$, 
A. MEDIAVILLA$^{3}$,
 J. V. TERAN$^{3}$, 
E. VILLA$^{3}$, 
L. PICCIRILLO$^{4}$, 
C. DICKINSON$^{4}$, 
K. GRAINGE$^{4}$, 
S. HARPER$^{4}$, 
M. McCULLOCH$^{4}$, 
S. MELHUISH$^{4}$, 
G. PISANO$^{4}$, 
R. A. WATSON$^{4}$,
A. LASENBY$^{5,9}$, 
M. ASHDOWN$^{5,9}$, 
Y. PERROTT$^{5}$, 
N. RAZAVI-GHODS$^{5}$, 
D. TITTERINGTON$^{5}$ and 
P. SCOTT$^{5}$ 
}

\address{$^{1}$ Instituto de Astrofis{\'{\i}}ca de Canarias, 38200 La Laguna, Tenerife, Canary Islands, Spain \\
$^{2}$ Instituto de F{\'{\i}}sica de Cantabria (CSIC-Universidad de Cantabria), Avda. de los Castros s/n, 39005 Santander, Spain \\
$^{3}$ Departamento de Ingenieria de COMunicaciones (DICOM), Laboratorios de I+D de Telecomunicaciones, 
Universidad de Cantabria, Plaza de la Ciencia s/n, E-39005 Santander, Spain \\
$^{4}$ Jodrell Bank Centre for Astrophysics, Alan Turing Building, 
School of Physics and Astronomy, The University of Manchester, Oxford
Road, Manchester, M13 9PL, U.K \\
$^{5}$ Astrophysics Group, Cavendish Laboratory, University of
Cambridge, J.J. Thomson Avenue, Cambridge CB3 0HE, UK \\
$^{6}$ Departamento de Astrof{\'{\i}}sica, Universidad de La Laguna
(ULL), 38206 La Laguna, Tenerife, Spain \\
$^{7}$ Consejo Superior de Investigaciones Cient{\'{\i}}ficas, Spain \\
$^{8}$ Departamento de F{\'{\i}}sica, Universidad de la Serena,
Av. Cisternas 1200, La Serena, Chile \\
$^{9}$ Kavli Institute for Cosmology, Madingley Road, Cambridge, CB30HA}

\maketitle\abstracts{
QUIJOTE (Q-U-I JOint TEnerife) is an 
experiment designed to achieve CMB B-mode polarization detection
and sensitive enough to detect a primordial gravitational-wave 
component if the B-mode amplitude is larger than $r$ = 0.05.
It consists in two telescopes and three intruments observing in the
frequency range 10-42 GHz installed at the Teide Observatory in the
Canary Islands, Spain. The observing strategy includes three raster
scan deep integration fields for cosmology, a nominal wide survey
covering the Northen Sky and specific raster scan deep integration 
observations in regions of specific interest. The main goals of the
project are presented and the first scientific results
obtained with the first instrument are reviewed.
}

\section{Introduction}

Over the last decades, astronomers have measured with unprecedented
precision many of the parameters that describe the current
Cosmological Model ~\cite{jarm1}~\cite{hin}~\cite{pla11}. 
Several experiments focus on the Cosmic Microwave Background (CMB) and
its anisotropies, that originated about 380.000 years after the Big
Bang. Analyses of the CMB suggest that, at a very early time, an
exponentially accelerated expansion of the Universe occurred. This
inflationary scenario ~\cite{gut} can solve most of the problems that haunted the classical
Big Bang scenario. A prediction from inflation is that during the
accelerated expansion, the quantum fluctuations in the dominant scalar
field would have grown into macroscopic density fluctuations (i.e.,
scalar fluctuations, which have been measured) and a background of
gravitational waves (GWB), (i.e., tensor fluctuations). Theoretical
models predict a B-mode whose amplitude is defined by the ratio of the 
tensor-to-scalar fluctuations and quantified with the parameter $r$,
constrains the expansion rate at that time providing a unique measure
of the energy scale of inflation. The GWB would have left a
characteristic imprint on the polarization of the CMB photons at last
scattering. This imprint can in principle be characterized by
estimating the parameter $r$ from the analysis of polarization maps. 
This is an extremely challeging multi-disciplinary task which requires 
very high sensitivity experiments and a full understanding of the 
polarization properties of the astrophysical signals emitted in our Galaxy and beyond.

\section{Prospects For CMB B-Mode Polarization Detection}

Several experiments investigate the temperature and polarization
properties of the CMB radiation. Until very recently, limits on the
B-mode amplitude came from the analysis of the angular power spectrum 
in temperature: Planck+WMAP+HighL (r$<$0.11 at the 95$\%$ 
Confidence Level or C.L.)~\cite{pla13-22}, SPT+WMAP+H0+BAO (r$<$0.17 at the 95$\%$ 
C.L.) and WMAP alone: r$<$0.36, at 95$\%$ C.L.~\cite{kei}. Improvement on such 
estimates have be obtained by the analysis of the CMB angular power 
spectrum in polarization. One constraint has come from the
BICEP/Keck and Planck collaboration with $r<$ 0.12 (95$\%$C.L.)~\cite{bic-pla}.
The $Planck$ collaboration alone reports an upper 
bound on the tensor-to-scalar ratio $r<$ 0.11 (95$\%$ C.L.)~\cite{pla20}.
The BICEP/Keck Array cosmic microwave background polarization 
experiments have recently reported the currently most stringent constraints on the 
tensor-to-scalar ratio to date with $r<0.09$ from B-modes alone, 
and $r<0.07$ in combination with other datasets~\cite{bic-keck}. The
POLARBEAR collaboration reports an improved measurements of the CMB
B-mode polarization power spectrum and rejects the null hypothesis of
no B-mode polarization detection at a confidence level of 3.1$\sigma$
over angular multipoles $500 \leq l \leq 2100$~\cite{pol}. 

The QUIJOTE~\cite{quijote10}~\cite{quijote17} (Q-U-I JOint TEnerife) is an 
experiment designed to achieve B-mode polarization detection
and sensitive enough to detect a primordial gravitational-wave 
component if the B-mode amplitude is larger than $r$ = 0.05. The other main 
science driver of this experiment is to characterize the polarization
of low-frequency foregrounds, mainly the synchrotron emission and 
the Anomalous Microwave Emission (AME), so that these signals can be 
removed from the primordial maps to a level that will permit reaching
the previous sensitivity on $r$ for cosmology. 

\section{The QUIJOTE Experiment}

The QUIJOTE experiment is a
collaboration between the Instituto de Astrof{\'{\i}}sica de Canarias,
the Instituto de F{\'{\i}}sica de Cantabria and DICOM University of Cantabria, in Spain, and the Universities of
Cambridge and Manchester, in the UK.
It consists of two telescopes and
three linear polarimetry instruments covering
respectively the frequencies 10-20, 31 and 42 GHz. The experiment is located at
Iza\~na, near the Teide Volcano, in the Tenerife island (Spain) at an
altitude of 2400 meters over the sea level at longitude, latitude position
$28.3^{\circ}$N, $16.5^{\circ}$W. The Iza\~na site is well suited for
such an experiment as it has been a test bed for previous CMB
experiments starting from the Tenerife experiment in 1984~\cite{tenerife}.

The development of the project includes two phases. In the first phase the
first QUIJOTE telescope (QT1) was installed and the
first two instruments were built. The first instrument, the
Multi-Frequency Instrument (MFI) observing in the frequency range
10-20 GHz had its first light on November
2012 and has been operating for almost 5 years now. In the current and
second phase the second telescope (QT2) that was installed on July 2014
has been tested and is operational. The second instrument, the 
Thirty-Gigahertz Instrument, or TGI, made of 30 receivers observing at
31 GHz has been built and already started its commisioning phase. The
third instrument, the Fourty-Gigahertz Instrument, or FGI, observing a frequency centred at 42
GHz, has been built and is currently in the integration phase.
Half of the pixels of the TGI and half of the pixels of the FGI are
currently being mounted on the same cryostat, to be installed at 
the QT2. The first light of this combination of
receivers and the commisioning phase of half of the FGI 
should start before end of 2017.

\subsection{The QUIJOTE Telescopes}

\begin{figure}
\centerline{\includegraphics[width=0.9\linewidth]{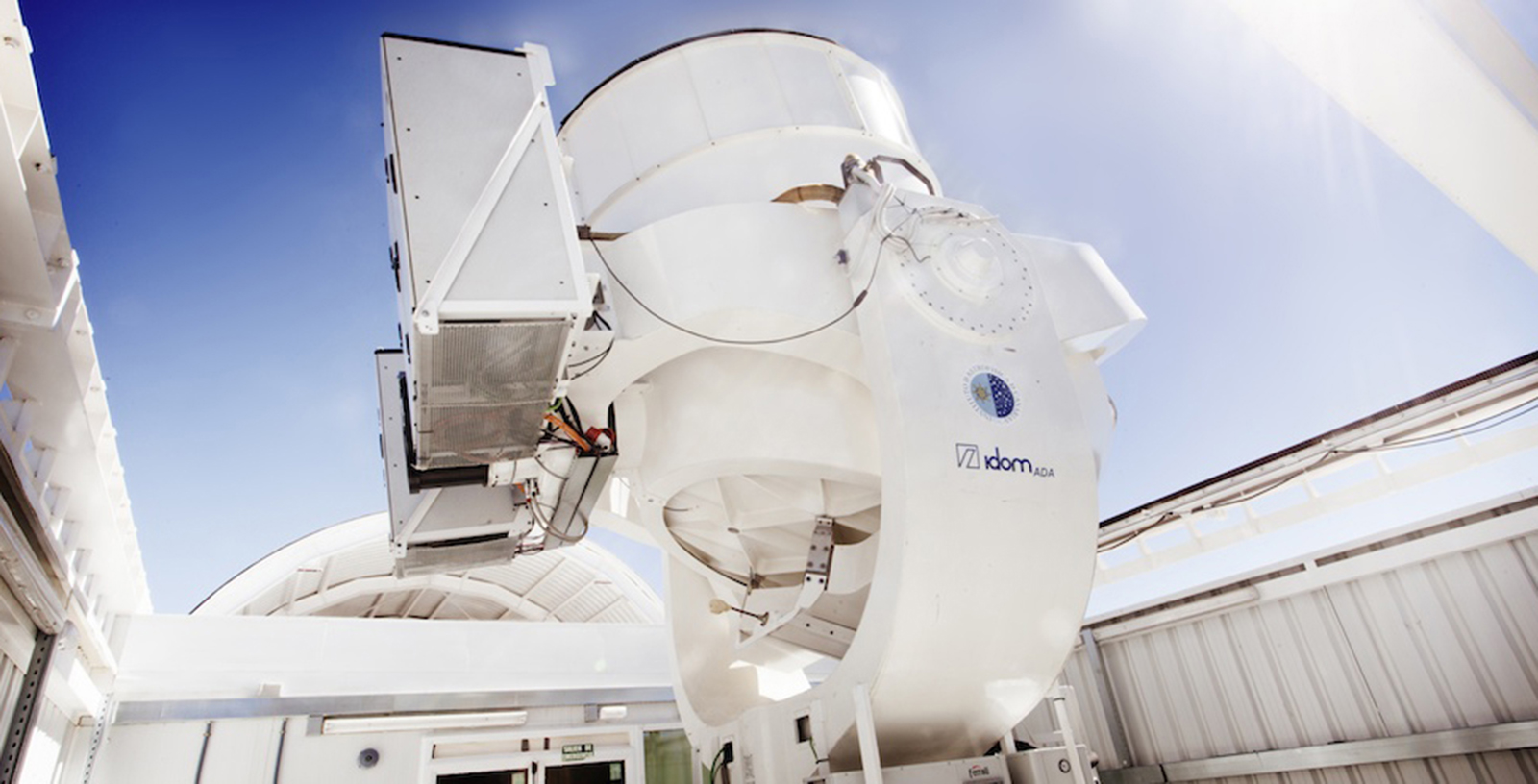}}
\caption[]{QUIJOTE telescope 1 in its enclosure at the Teide Observatory.}
\label{fig:qt1}
\end{figure}

The two QUIJOTE teslecopes, QT1 (see Figure~\ref{fig:qt1}) 
and QT2, have altazimuthal mounts with 
crossed-dragoned designed. The primary (parabolic) and 
secondary (hyperbolic) mirrors have
apertures of 2.25 m and 1.90 m, respectively. Each telescope has a
maximum azimuthal speed of 0.25 Hz or 15 rpm and can point at a
minimum elevation of 30$^{\circ}$. The data obtained from the QT1 with
the MFI show a high symmetric beam with ellipticity $> 0.98$ with very low far
sidelobes ($\leq -40$dB) and polarization leakage ($\leq -25$dB).

\subsection{The Multi-Frequency Instrument: MFI}

\begin{figure}
\centerline{\includegraphics[width=0.6\linewidth]{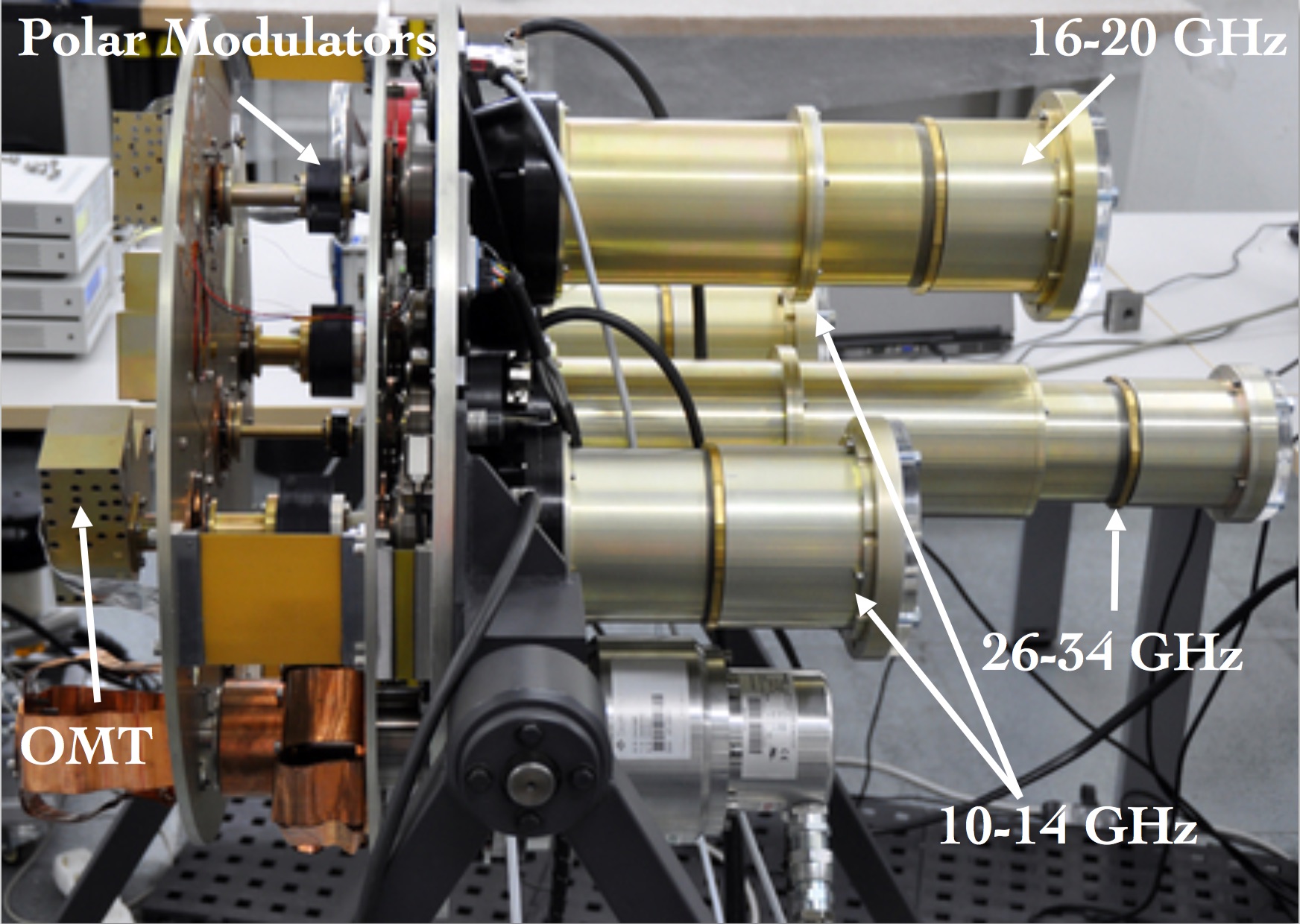}}
\caption[]{MFI Instrument.}
\label{fig:mfi}
\end{figure}

The first instrument of the QUIJOTE experiment is the Multi-Frequency Instrument, or
MFI~\cite{mfi}. It is operating in four frequency bands centred at 11.2, 12.9,
16.7 and 18.7 GHz and is operative from November 2012. It is made of
four horns, two of them operating in the frequency range 10-14 GHz and
the two other ones operating in the frequency range 16-20 GHz.
The modulation of the polarization is operated by a mechanically
rotating half-wave plate. The sensitivities are of order 400-600
$\mu$K$/s^{1/2}$ per channel. Each horn has 4 channels operating at one of
the two frequencies and their  
combinations permits to recover the I, Q and U Stokes parameters at
each frequency. The FWHMs are in the range 0.62$^{\circ}$ - 0.87$^{\circ}$.
Since its first light the MFI has been performing routine observations 
covering large sky areas ~\cite{quijote10}. Some parts of the instrument are shown in Figure~\ref{fig:mfi}.

\subsection{The Thirty and the Fourty GHz Instruments: TGI and FGI}

The TGI, is a detector including 30
pixels centred at a frequency of 31 GHz with a bandwidth of 10 GHz. 
Thirty similar feedhorns or pixels have been designed with linear
polarimetry capability. Each pixels is made of 4 channels providing
data to the acquisition system. The polarization modulation is obtained by combining two
phase-switches, each of them having two different possible phase
states, i.e. $0^{\circ}/90^{\circ}$ and $0^{\circ}/180^{\circ}$,
respectively. High frequency modulation allows to get almost
simultaneous measurements of $I$, $Q$ and $U$ on the sky 
and to get rid of many systematics. The FGI, has a design similar to the one of the TGI with a center
frequency of 41 GHz in a bandwidth of 12 GHz and the two instruments
can share a common cryostat. The acquisition system has been designed
to be operational with the two type of pixels. 
A schematic view of the TGI receiver is shown in Figure~\ref{fig:mfi}.

\begin{figure}
\centerline{\includegraphics[width=0.9\linewidth]{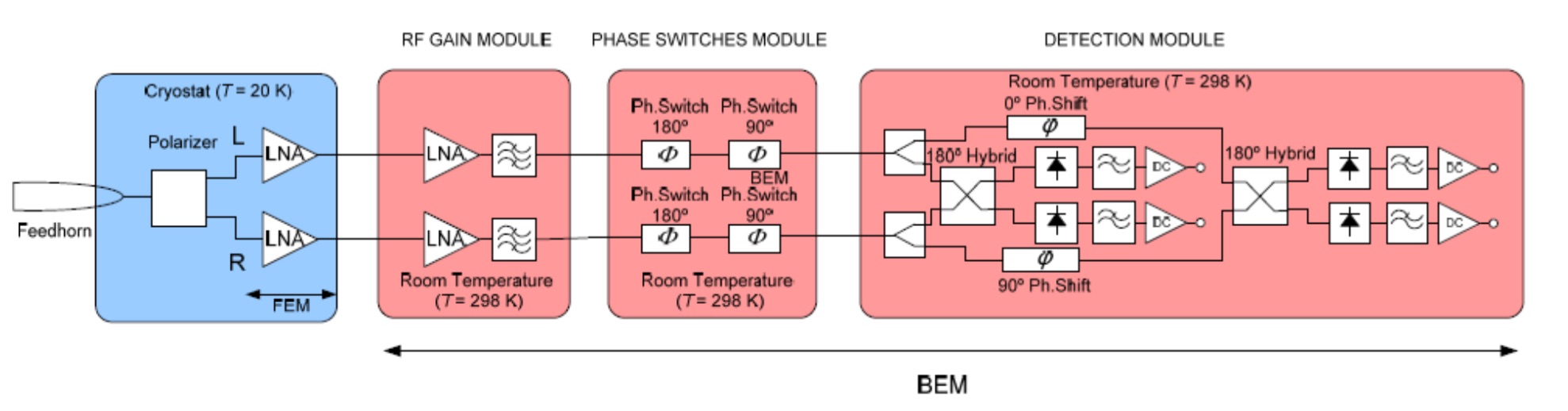}}
\caption[]{Schematic view of the TGI.}
\label{fig:tgi}
\end{figure}

\section{Observation strategy}

\begin{figure}
\centerline{\includegraphics[width=0.9\linewidth]{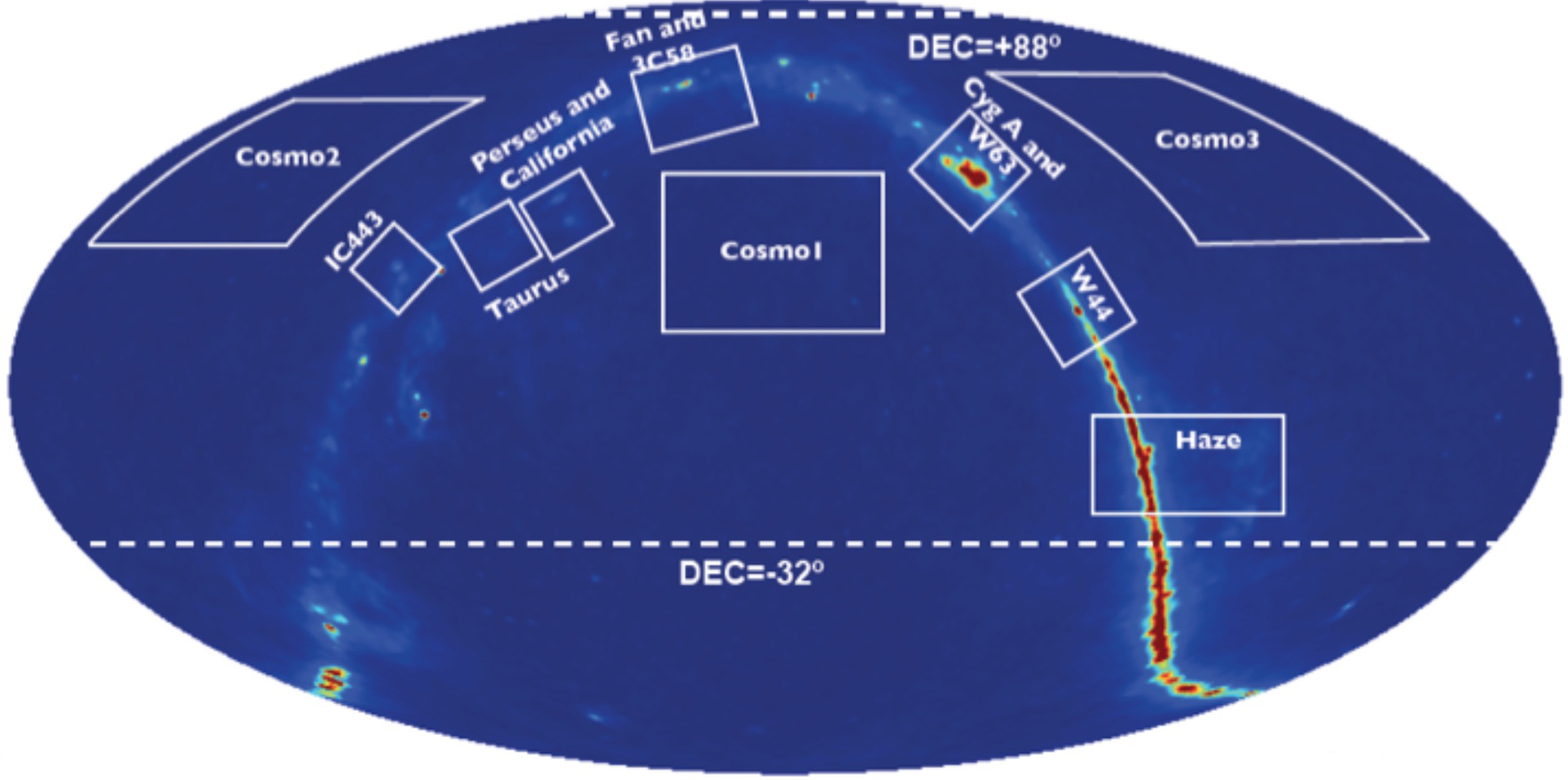}}
\caption[]{QUIJOTE fields of view displayed over the WMAP 22.7 GHz map
in equatorial coordinates.}
\label{fig:quijote_fields}
\end{figure}

The QUIJOTE experiment will take advantage of the combination and
sensitivity of its three instruments to characterize the nature of the CMB emission in
intensity and in polarization toward 3 fields dedicated to cosmology.
The data obtained with the MFI, TGI and FGI will be combined with 
the Planck and WMAP data and allow a state-of-the-art 
characterization of the foregrounds between 10 and 42 GHz to produce
CMB maps at 31 GHz and 42 GHz. For this purpose the role of the MFI is fundamental in the
sense that it will allow to disentangle the contributions of the
Free-Free, of the AME and Synchrotron components from the thermal dust 
component in the frequency range 10-20 GHz. The core science program includes
observations toward three fields dedicated to cosmology and a wide
survey. Additional fields out of the core science program are
dedicated to targets of specific science interest.
A view of the sky covered from the Teide Observatory including the
wide survey, the 3 cosmology fields and some of the other  
targets of interest is given in Figure~\ref{fig:quijote_fields}.

\subsection{Cosmology Fields}

The observing strategy includes deep integration fields obtained with raster scan 
mode toward three fields dedicated to cosmology. The three fields
cover a total of about 3000 deg$^{2}$. From its first light in
November 2012 a total of about 4000 hours of observations have been
obtained with the MFI toward those three fields. Given the expected nominal 
sensitivities to be reach by the TGI and the FGI, a
sensitivity on the tensor-to-scalar ratio of $r=0.1$ (at 95$\%$ C.L.)
should be obtained after one effective year of observations with the TGI
toward the total area of the three cosmology fields. A sensitivity of
$r$ of 0.05 (at 95$\%$ C.L.) will necessitate the combination of 3
effective years of observations with the FGI and 2 effective years of
observations with the FGI.

\subsection{Wide survey and Galactic Foregrounds Characterization}

One of the main science driver of the QUIJOTE experiment with the
MFI is to characterize the properties of the synchrotron emission,
i.e. the large scale magnetic field, spectral index, curvature of the
index and polarization properties, and to characterize the properties
of AME which may be a polarization CMB contaminant.
In order to do so the observing strategy includes a wide survey of
about 20,000 deg$^2$
obtained from observations in nominal mode, i.e., observations
obtained over blocks of about 24 hours at constant elevations (EL=
30$^{\circ}$, 40$^{\circ}$, 50$^{\circ}$, 60$^{\circ}$, 65$^{\circ}$,
70$^{\circ}$, 75$^{\circ}$ and 80$^{\circ}$) at a scanning
speed of 2 rpm that have accumulated a total of about 6800 hours.

\begin{figure}
\centerline{\includegraphics[width=0.9\linewidth]{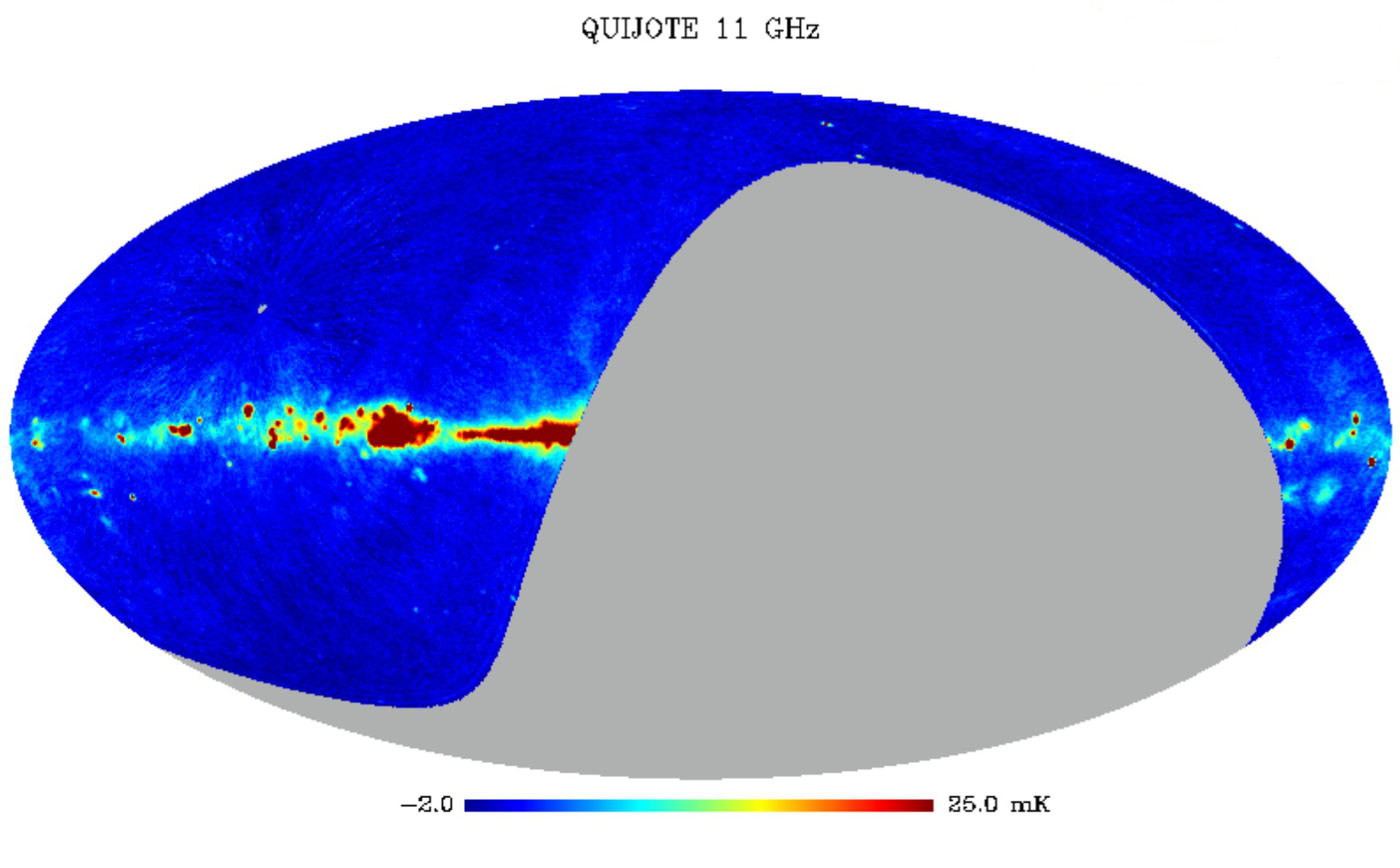}}
\caption[]{Preliminary map at 11 GHz of the QUIJOTE wide survey.}
\label{fig:preliminary_map}
\end{figure}

The wide survey is part of the Radioforeground Project \footnote{ \url{http://www.radioforegrounds.eu}}, one of the
H2020-COMPET-2015 selected project within the European Union's 
Horizon 2020 research and innovation programme. In addition to the 
physics results and new knowledge it is going to produce (as mentionned above), it is planned to provide  
a complete and statistically significant multi-frequency catalogue of radio sources in
both temperature and polarization as well as specific (open source)
software tools for data processing, data visualization and public information.
A preliminary wide survey map obtained with a total of 700 hours of
observations at 11 GHz with the MFI is shown
in Figure~\ref{fig:preliminary_map}.

\subsection{Observations dedicated to specific regions of interest}

A series of specific fields of interest has been defined which are
observed in raster scan mode in order to complement the wide survey 
with deeper integrations. These fields 
include Galactic region like the Fan region, a region toward which a large
uniform magnetic field is probed which origin is still not understood~\cite{hill17}. 
A series of molecular clouds that have been studied by the Planck
collaboration~\cite{pla15} and are sources where the AME emissivity
was detected with high significance. This observation program includes
the Taurus molecular cloud and is part of the
PolAME project \footnote{ \url{http://www.iac.es/proyecto/polame/}} dedicated to the study of AME properies in cold
molecular clouds supported by the Marie Sklodowska-Curie 
European Union's Horizon 2020 research and innovation programme.  
Other regions of scientific interest are the Galactic center and the
Haze regions, 3C58 and M31. The Perseus molecular cloud complex
region, the W43 and W47 molecular cloud regions as well as the Super
Novae Remnants (SNRs) W44, W51, IC443 and W63 have also been observed.  

\section{First Scientific Results}

\subsection{The Perseus Molecular Cloud Complex}

The first scientific results obtained with the MFI are from the
analysis of 194 hours of observations towards the well studied Perseus
molecular cloud complex~\cite{qsr1} over an area of about 250
deg$^{2}$. The flux densities obtained
with the MFI in the frequency range 10-20 GHz nicely complete the WMAP
measurements and allow a full characterization of the AME component
which would not be possible otherwise.  
The upper limits on the fraction of polarization of 
the total intensity components derived from the Spectral Energy
Distributions (SEDs) analysis of G159.6-18.5 are $\pi <
6.3\%$ and $\pi < 2.8 \%$ at 12 and 18 GHz, respectively, whichs means
upper limits $\pi_{\rm AME} <10.1 \% $ and $\pi_{\rm AME} < 3.4 \%$
on the fraction of polarization associated with the AME components. 

\subsection{The W43, W44 and W47 SNRs}

A total of 210 hours observations have been obtained along the Galactic
Plane toward the area 24$^{\circ} < l < 45^{\circ}$, $\mid b \mid < 8^{\circ}$. 
Combined with the WMAP data the frequency range covered by the MFI is crucial to confirm the presence
of AME towards the two molecular cloud complexes W43 (at 22$\sigma$) and
W47 (at 8 $\sigma$). The most stringent constraints ever obtained on the polarisation fraction of the AME
are obtained from the analysis of the polarised flux of W43 and show
$\pi_{\rm AME} < 0.39\%$ (95 per cent C.L.) from the QUIJOTE 17 GHz data, and $\pi_{\rm AME} < 0.22\%$ from WMAP 41 GHz data.
The estimated spectral index of the synchrotron emission in W44 is,
$\beta_{\rm sync} =−0.62 \pm 0.03$, in good agreement with the value inferred from the intensity
spectrum once a free$-$free component is included in the fit.
The change in the polarisation angle associated with Faraday rotation
in the direction of W44 corresponds to a rotation measure of $−404 \pm
49 $ rad /m$^{2}$.

\subsection{The Taurus Molecular Cloud}

\begin{figure}
\centerline{\includegraphics[width=1.\linewidth]{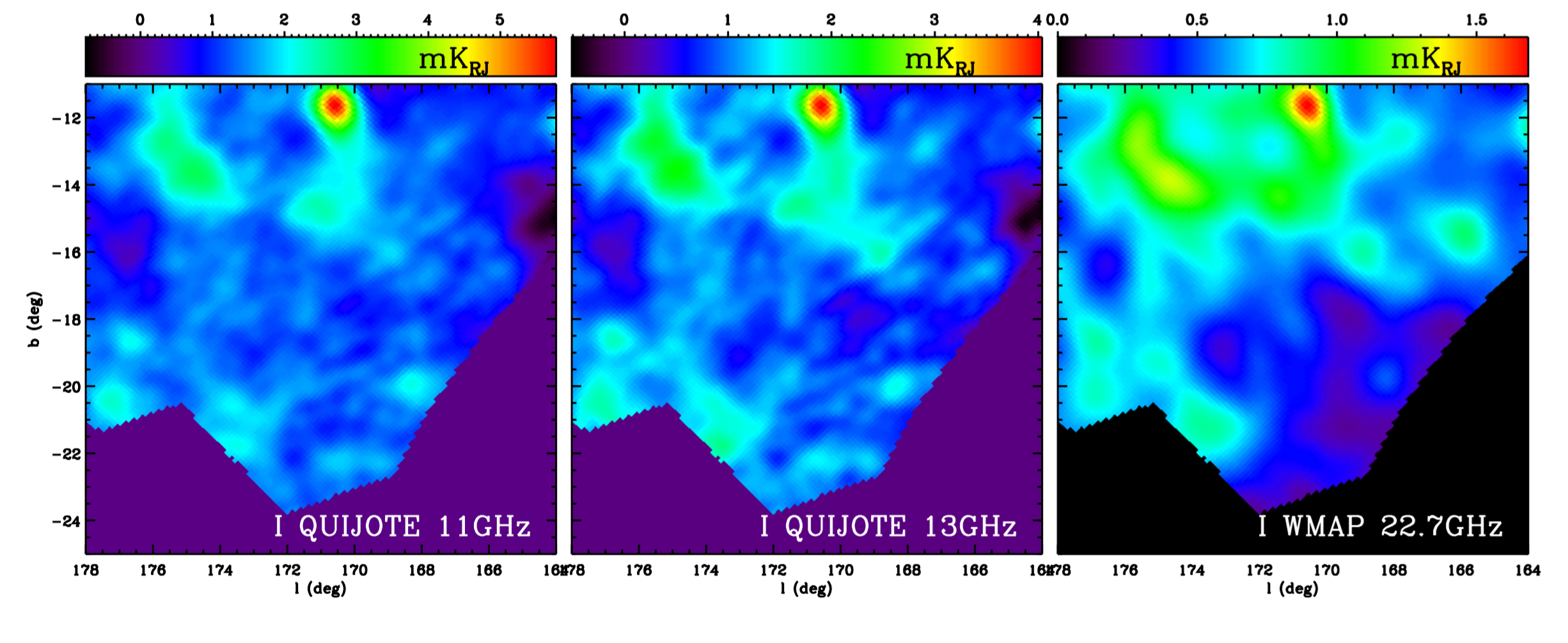}}
\caption[]{Left: TMC map at 11 GHz
  obtained with the MFI on QT1. Middle: same as left but at a
  frequency of 13 GHz. Right: same as left
  and middle but at a frequency of 22.7 GHz as obtained by the WMAP.}
\label{fig:tmc}
\end{figure}

A total observing time of about 423 hours has been obtained towards
the Taurus Molecular Cloud (TMC). The maps obtained at 11 GHz and 13
GHz shown in Figure~\ref{fig:tmc}  can be compared to the WMAP map at 22.7 GHz.
The analysis of the SED on the high column density regions of the TMC
shows a clear detection of AME in the TMC in the frequency range 10-60 GHz.
These results are presented and discussed in~\cite{poi18}.

\section{Conclusions}

The QUIJOTE is a fruitfull experiment starting to provide unique
measurements of the Nothern Hemisphere Sky in the frequency range 10-20 GHz with
its first instrument the MFI. The results provided by the MFI are strategical
to understand the properties of the free-free, AME and of the
synchrotron radiation so that they can be properly
removed from the CMB polarization map that are going to be provided by
the TGI and FGI. Given the expected nominal 
sensitivities to be reach by these two instruments, a
sensitivity on the tensor-to-scalar ratio of $r=0.05$ (at 95$\%$ C.L.)
wil need the combination of 3
effective years of observations with the FGI and 2 effective years of
observations with the FGI. 

\section*{Acknowledgments}

The QUIJOTE experiment is being developed by the Instituto de
Astrof{\'{\i}}sica de Canarias (IAC), the Instituto de F{\'{\i}}sica de Cantabria
(IFCA), and the Universities of Cantabria, Manchester and Cambridge. 
Partial financial support is provided by the Spanish Ministry of
Economy and Competitiveness (MINECO) under the projects 
AYA2007-68058-C03-01, AYA2010-21766-C03-02, AYA2014-60438-P, 
and also by the Consolider-Ingenio project CSD2010-00064 
(EPI: Exploring the Physics of Inflation). This project has received
funding from the European Union's Horizon 2020 research and 
innovation programme under grant agreement number 687312 (RADIOFOREGROUNDS)
and number 658499 (PolAME). FP is a Marie Sklodowska-Curie fellow from the 
European Union's Horizon 2020 research and innovation programme 
under grant agreement number 658499 (PolAME).

\end{document}